\documentclass[fleqn,usenatbib]{mnras}

\usepackage{newtxtext,newtxmath}

\usepackage[T1]{fontenc}

\usepackage{booktabs}
\usepackage{multirow} 
\usepackage{graphicx}

\usepackage{multicol}	
\usepackage[english]{babel}
\usepackage{epstopdf}
\usepackage{ae,aecompl}
\usepackage{hyperref}
\usepackage{amsmath}    

\def\msun{\,h^{-1}{\rm M}_\odot}

\usepackage{color}

\makeatletter

\newcommand{\Rmnum}[1]{\expandafter\@slowromancap\romannumeral #1@}
\makeatother

\title[Robustness of cosmic void statistics]{Robustness of cosmic void statistics: insights from SDSS DR7 and the ELUCID simulation}

\author[Youcai Zhang]{Youcai Zhang$^{1}$\thanks{yczhang@shao.ac.cn},
Xiaohu Yang$^{2,3}$, Hong Guo$^{1}$, Peng Wang$^{1}$, 
Feng Shi$^{4}$
\\
$^{1}${Shanghai Astronomical Observatory, Nandan Road 80, Shanghai 200030,
  China} \\
$^{2}$State Key Laboratory of Dark Matter Physics, Tsung-Dao Lee Institute \& School of Physics and Astronomy, Shanghai Jiao Tong University, \\~~~Shanghai 201210, China\\
$^{3}$Shanghai Key Laboratory for Particle Physics and Cosmology, and Key Laboratory for Particle Physics, Astrophysics and Cosmology, \\~~~Ministry of Education, Shanghai Jiao Tong University, Shanghai 200240, China\\
$^{4}$School of Aerospace Science and Technology, Xidian University, Xi'an 710126, China
}

\begin{document}
\label{firstpage}
\pagerange{\pageref{firstpage}--\pageref{lastpage}}
\maketitle

\begin{abstract}
We present a systematic analysis of the statistical properties of cosmic voids using galaxies from the Sloan Digital Sky Survey Data Release 7 (SDSS DR7) and subhaloes from the ELUCID constrained simulation. By comparing voids identified in redshift space, real space, and reconstructed volumes, we assess the impact of redshift-space distortions (RSD) and tracer bias. Using the \texttt{VAST} toolkit, we apply both the geometry-based \texttt{VoidFinder} algorithm and watershed-based methods. We find that void properties are not equally robust. The three-dimensional morphology of voids, quantified by their sphericity and triaxiality, remains stable across different reconstructions and tracer selections. In contrast, void size distributions and radial density profiles depend strongly on the identification algorithm, with watershed-based methods systematically producing larger voids and higher compensation walls than \texttt{VoidFinder}. Using the full ELUCID simulation box, we show that tracer bias mainly affects void density profiles, with noticeable changes only for the most massive subhaloes ($>10^{11.5}\msun$). The agreement between SDSS observations, the ELUCID reconstruction, and the full simulation box demonstrates the high fidelity of constrained simulations and reveals a clear hierarchy in the robustness of void statistics.
\end{abstract}


\section{Introduction}\label{sec_intro}

The large-scale structure of the Universe is characterized by a complex network known as the cosmic web, consisting of dense clusters, elongated filaments, expansive sheets, and vast underdense regions called cosmic voids \citep{Bond1996}. 
As the most voluminous components of the cosmic web, voids represent relatively pristine environments for studying structure formation. 
Their interiors are generally thought to be less influenced by the complex non-linear baryonic processes that dominate high-density regions.
Owing to their low-density interiors and relatively simple dynamical evolution, cosmic voids have emerged as a powerful laboratory for testing cosmological models and modified gravity \citep{Takadera2025, Banik2026, Moretti2026a, Moretti2026b, Moretti2026c}, probing galaxy bias \citep{Alfaro2026}, and elucidating the connection between galaxy evolution and the underlying dark matter distribution \citep{Rodrguez2024, Azevedo2026, Fuad2026, London2026, Rouse2026}.

Over the past two decades, numerous studies have investigated the statistical properties of voids in both cosmological simulations and large galaxy surveys \citep{Tavasoli2026, Xu2026}. Void statistics, such as the size distribution \citep{Contarini2019, Ronconi2019, Verza2019}, shape properties \citep{Platen2008, Lavaux2010, Lavaux2012, Bos2012, Ebrahimi2024}, and radial density profiles \citep{Hamaus2014}, have been shown to encode valuable information about the formation and evolution of large-scale structure. 

Theoretically, voids originate from the gravitational amplification of primordial density fluctuations, 
as described in the excursion-set framework for void evolution \citep{Sheth2004}.
In numerical simulations, this evolutionary process can be precisely tracked from the initial conditions to the present day, revealing how voids are effectively evacuated as matter flows toward the surrounding structure. Complementarily, in observations, large galaxy redshift surveys have enabled the identification of extensive void catalogues across different cosmic epochs \citep{Mao2017, Douglass2023, Rincon2025, Arsenov2026}, allowing for detailed statistical comparisons between these theoretical predictions and observational measurements.

Despite these extensive efforts, it remains unclear whether the statistical properties of cosmic voids are intrinsically robust, or to what extent they are shaped by observational systematics and algorithmic definitions. Therefore, extracting cosmological information from voids depends crucially on a precise understanding of how their statistical properties are affected by observational systematics. This task is complicated by several layers of systematic effects. 

First, the definition and identification of voids are not unique; different algorithms partition the cosmic web differently \citep{Platen2007}, leading to catalogues with varying size distributions and morphologies. To facilitate systematic comparisons between different void definitions, integrated analysis toolkits such as the \texttt{Void Analysis Software Toolkit} (\texttt{VAST}; \citealp{Douglass2023}) have recently been developed. \texttt{VAST} provides a common computational framework that enables the application of both geometry-based sphere-finders \citep[e.g. \texttt{VoidFinder},][]{Hoyle2002} and watershed-based density-tracing algorithms \citep{Neyrinck2008, Sutter2015, Nadathur2019a} within a consistent analysis environment.

Second, observational effects such as Redshift-Space Distortions (RSD) play a critical role in shaping the observed void morphology. Induced by the peculiar velocities of tracer galaxies, RSD introduce prominent anisotropic signatures in both void density profiles and their geometric shapes, potentially causing voids to appear elongated or flattened along the line of sight \citep{Hamaus2015, Cai2016, Nadathur2019c}. To mitigate these nonlinear effects and restore the underlying real-space clustering, cosmological reconstruction has become an essential tool to improve the precision of geometric constraints \citep{Nadathur2019b}. However, it remains a critical question whether these reconstruction techniques can accurately recover the intrinsic, isotropic geometric shapes of voids, especially given the sparse tracer density within their interiors.

Finally, beyond these kinematic distortions, the inherent tracer bias further modulates the connection between galaxies and the underlying matter distribution \citep{Schuster2025}. Consequently, these tracer-dependent signatures may bias the recovered void geometry, even when the observational data are mapped back to real space.

Tackling these challenges has inspired two complementary strategies in recent research. One strategy concentrates on enhancing the physical robustness of void-finding algorithms. 
Examples include dynamical methods such as the Algorithm for Void Identification in coSMology (\texttt{AVISM}; \citealp{Monllor2025}), which incorporates velocity divergence to enforce dynamical consistency; the Back-In-Time Void Finder (\texttt{BitVF}), which reconstructs Lagrangian dynamics using optimal transport \citep{Sartori2026}; and recent probabilistic frameworks that employ deep graph neural networks and flow matching for Bayesian void inference \citep{Tavasoli2026}. While these methods aim to mitigate tracer bias and observational distortions through more sophisticated modeling, their performance inevitably depends on the adopted dynamical assumptions or the training priors of the underlying models.

An alternative strategy is to improve the fidelity of the underlying data-simulation 
comparison framework. In particular, constrained simulations that reproduce the 
observed large-scale structure offer a powerful platform for directly linking galaxy 
observations to the underlying dark matter distribution. Such simulations enable controlled tests of observational systematics while preserving the actual large-scale structure observed in the local Universe.

In this study, we leverage the constrained ELUCID (\textit{Exploring the Local Universe with the reConstructed Initial Density field}) simulation to provide a unique, \textit{apples-to-apples} comparison platform \citep{Yang2007, Yang2012, WangHuiyuan2012, WangHuiyuan2014, WangHuiyuan2016, WangHuiyuan2018}. By using observed galaxy groups from the Sloan Digital Sky Survey Data Release 7 (SDSS DR7) to reconstruct the initial conditions, ELUCID enables a direct simulation of the \textit{particular} large-scale structure observed in the local Universe. This constrained framework allows us to link observed galaxies directly to their underlying dark matter haloes within a simulated cosmic web that broadly reproduces the large-scale topology and geometry of the SDSS volume \citep{Yang2018}.

Using the \texttt{VAST} package \citep{Douglass2023}, we identify void populations across multiple datasets, including the SDSS DR7 galaxy sample and subhaloes from the ELUCID simulation. The unified implementation of multiple algorithms within \texttt{VAST} enables a consistent cross-check of how void properties depend on identification methodology. Our analysis focuses on three fundamental statistical properties: the size distribution (effective radius), three-dimensional morphology (sphericity and triaxiality), and stacked radial density profiles.

The primary objective of this work is to systematically evaluate the robustness of these void statistics against the main systematic hurdles: RSD and tracer bias. 
To achieve this, our analysis is structured into three key steps. 
First, by comparing voids identified in SDSS redshift space, reconstructed SDSS real space, and the ELUCID constrained volume, we investigate how void statistics respond to RSD mitigation and reconstruction procedures. Second, we verify the statistical representativeness of the constrained ELUCID volume by comparing it against the full simulation box. Finally, we exploit the larger volume of the periodic simulation box to investigate how void properties vary with subhalo tracer mass, thereby quantifying the imprints of tracer bias. This comprehensive validation helps assess the applicability and limitations of constrained simulations for studying the low-density Universe in the era of next-generation surveys.

This paper is organized as follows. Section~\ref{sec_data} describes the observational data from SDSS DR7, the ELUCID constrained simulation, and the void-finding algorithms employed. Section~\ref{sec:results} presents our main results: the void size distribution, a detailed analysis of void shapes, and the stacked density profiles. Within each subsection, we systematically compare observations with simulations and examine dependencies on algorithm and tracer mass. Finally, Section~\ref{sec_summary} summarizes our key findings and discusses their implications.

\section{Data and Method}\label{sec_data}

\subsection{Observational data}
\label{obs_data}

\begin{figure}
\centering
\includegraphics[width=0.45\textwidth]{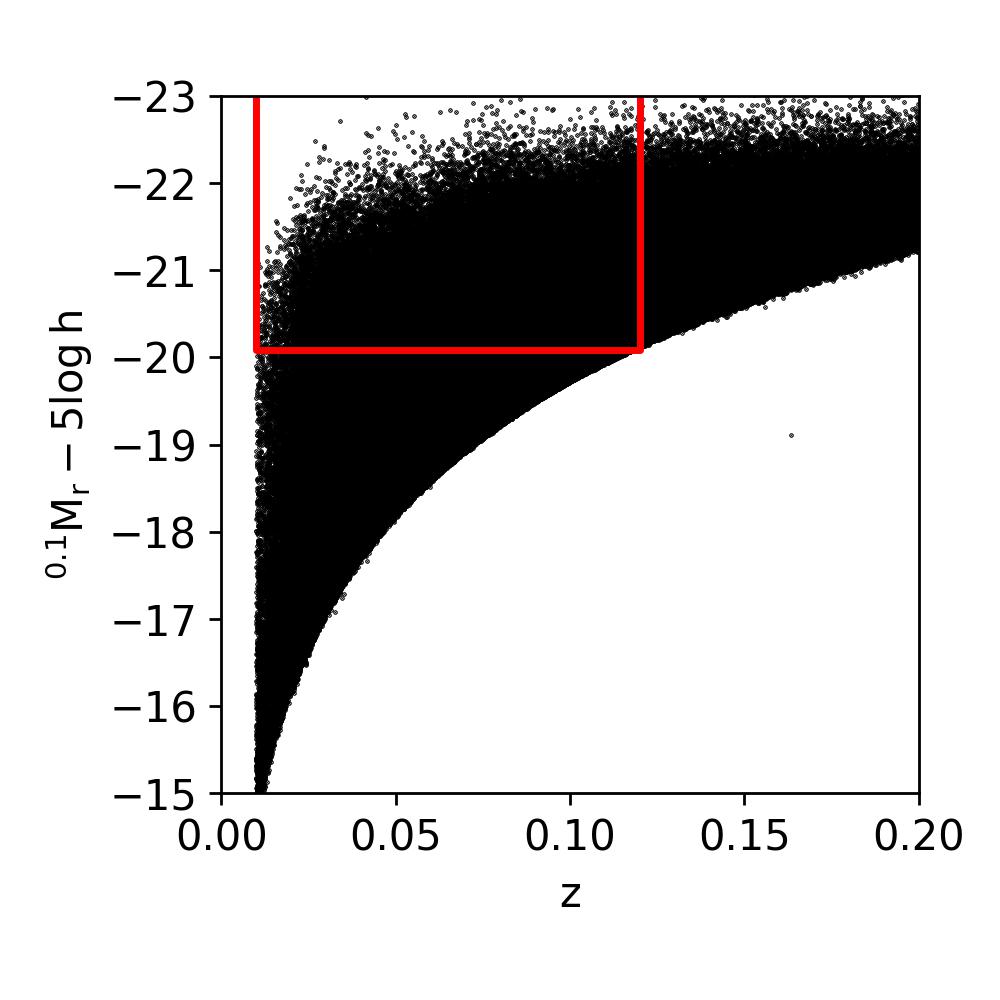}
\vspace{-0.5cm}
\caption{Illustration of the volume-limited sample selection for SDSS galaxies. The rectangular region in the redshift--absolute magnitude plane defines the sample boundaries: $0.01 \le z \le 0.12$ and $^{0.1}M_r - 5 \log h < -20.09$. The vertical line at $z=0.12$ represents the redshift completeness limit corresponding to the chosen luminosity threshold.}
\label{fig:slice_sdss}
\end{figure}

This study utilizes a sample of 639,359 galaxies selected from the New York University Value-Added Galaxy catalogue \citep[NYU-VAGC;][]{Blanton2005}, based on SDSS DR7 data \citep{York2000, Abazajian2009}. Following the methodology of \citet{Shi2016}, we focus on the major continuous region in the Northern Galactic Cap (NGC), confined to celestial coordinates of $111^{\circ} < \alpha < 264^{\circ}$ and $-3^{\circ} < \delta < 68^{\circ}$. Within this footprint, a volume-limited sample is defined by applying a redshift range of $0.01 \le z \le 0.12$ and an $r$-band absolute magnitude limit of $^{0.1}M_r - 5 \log h < -20.09$, as illustrated in Figure~\ref{fig:slice_sdss}. These magnitudes are K-corrected and evolution-corrected to $z=0.1$ \citep{Blanton2003a, Blanton2003b}. 
This selection yields 158,840 galaxies in real space, corrected for RSD according to the method of \citet{Shi2016}, compared to 158,051 galaxies in the original redshift space. The $0.5\%$ discrepancy in galaxy counts arises from individual objects shifting across the survey boundaries during the RSD correction process. 

To briefly summarize the reconstruction pipeline, the identified galaxy groups are adopted as observational proxies for the underlying dark matter haloes \citep{Yang2007}. The total peculiar velocity of a galaxy is decomposed into the bulk velocity of its host halo center (associated with the large-scale Kaiser effect) and the internal virialized random motion within the halo (associated with the small-scale Fingers-of-God; FoG effect). The large-scale Kaiser distortion is then approximately corrected by reconstructing the bulk velocity field $\boldsymbol{v}_{\mathrm{cen}}(\boldsymbol{k})$ from the halo overdensity field $\delta_{\mathrm h}(\boldsymbol{k})$ in Fourier space via: 
\begin{equation}
\label{eqn:vel_cen}
\boldsymbol{v}_{\mathrm{cen}}(\boldsymbol{k}) = H a f(\Omega) \frac{i\boldsymbol{k}}{k^2} \frac{\delta_{\mathrm h}(\boldsymbol{k})}{b_{\mathrm{hm}}},
\end{equation}
where $H$ is the Hubble parameter, $a$ is the cosmic scale factor, $f(\Omega) \equiv \mathrm{d}\ln D/\mathrm{d}\ln a$ denotes the linear growth rate, and $b_{\mathrm{hm}}$ is the linear halo bias parameter. Using the line-of-sight component of the reconstructed bulk velocity ($v_{\mathrm{cen, los}}$), the corrected cosmological redshift $z_{\mathrm{corr}}$ for each group is iteratively updated until convergence according to:
\begin{equation}
\label{eqn:z_corr}
z_{\mathrm{corr}} = \frac{z_{\mathrm{obs}} - v_{\mathrm{cen, los}}/c}{1 + v_{\mathrm{cen, los}}/c},
\end{equation}
where $z_{\mathrm{obs}}$ denotes the observed redshift of the group, defined as the luminosity-weighted average redshift of its member galaxies. Subsequently, the small-scale FoG effect is statistically mitigated by compressing satellite galaxies along the line of sight toward the group center, assuming a Navarro-Frenk-White (NFW) profile associated with the host halo mass. 

Comprehensive mock catalogue tests presented in \citet{Shi2016} suggest that this reconstruction pipeline is able to recover the large-scale real-space clustering with relatively small residual systematic uncertainties. In particular, over scales of $0.2 \, h^{-1}\mathrm{Mpc} \le r \le 20 \, h^{-1}\mathrm{Mpc}$, the average systematic deviation in the reconstructed two-point correlation function (2PCF) remains below approximately $5\%$, while the quadrupole component is substantially suppressed relative to the original redshift-space sample. These tests indicate that the reconstructed catalogue provides a useful approximation to the underlying real-space distribution of galaxies and groups on the scales relevant to the present analysis, thereby offering a useful comparative framework for assessing how void statistics respond to RSD mitigation and reconstruction-related systematics.

This NGC region and redshift interval are identical to those used to reconstruct the initial conditions of the ELUCID constrained simulation \citep{WangHuiyuan2016}, which utilizes galaxy groups identified via the \citet{Yang2007} algorithm. For our comparative analysis, we adopt a subhalo mass threshold of $10^{11.8} \msun$ to match the number density of the observational sample, ensuring a robust statistical benchmark for void identification.

\subsection{Numerical simulation: ELUCID}
\label{sec:elucid_sim}

In this work, we utilize the ELUCID dark matter-only constrained simulation, which is designed to reproduce the large-scale structure of the local Universe using observational constraints derived from SDSS galaxy groups \citep{WangHuiyuan2014, WangHuiyuan2016}. The simulation provides a dynamically reconstructed realization of the nearby cosmic web and serves as a comparative real-space reference for our void analysis.

For clarity, the relationship between the observational reconstruction and the constrained realization procedure is schematically summarized in Equation~\ref{eqn:workflow} as a strictly sequential pipeline, where each descending step takes the output from the right-hand side of the preceding line as its direct input (see also \citealt{WangHuiyuan2016}):
\begin{equation}
\label{eqn:workflow}
\begin{aligned}
& \text{SDSS in Redshift Space} \\
& \quad \xrightarrow{\text{Approx.\ RSD Removal}} \text{SDSS in Real Space} \\
& \quad\quad \xrightarrow{\text{Halo-domain}} \text{Real-space Density Field} \\
& \quad\quad\quad \xrightarrow{\text{HMC+PM Optimization}} \text{Initial Conditions} \\
& \quad\quad\quad\quad \xrightarrow{\text{\textit{N}-body Evolution}} \text{ELUCID Realization}
\end{aligned}
\end{equation}
The reconstruction constraints are derived from SDSS galaxy groups with masses larger than
$10^{12} \msun$ within the survey volume
($0.01 \le z \le 0.12$), identified using the halo-based group finder of \citet{Yang2007, Yang2012}.
The observational galaxy-group catalogue first undergoes an approximate reconstruction procedure following \citet{WangHuiyuan2012}, in which RSDs are statistically suppressed to obtain an approximate real-space catalogue.
This density-velocity inversion procedure is methodologically identical to the large-scale reconstruction framework described in Section~\ref{obs_data} (Equations~\ref{eqn:vel_cen} and~\ref{eqn:z_corr}), since the \citet{Shi2016} pipeline directly extends the linear reconstruction formalism of \citet{WangHuiyuan2012} by incorporating a statistical mitigation of the small-scale FoG effect within the host groups. 

Based on this reconstructed real-space catalogue, the halo-domain reconstruction method developed by \citet{WangHuiyuan2009, WangHuiyuan2013} is applied to infer the corresponding mass density field.
The ELUCID constrained realization is then generated using a Hamiltonian Markov Chain (HMC) framework combined with approximate particle-mesh (PM) dynamics to infer the initial conditions whose evolved density field reproduces the reconstructed large-scale structure derived from the SDSS catalogue.
Within this framework, the forward modeling describes the physical evolution of matter in real space, while RSDs are treated separately as observational effects associated with line-of-sight peculiar velocities.
Large-scale Fourier modes are constrained by the observational reconstruction, whereas unconstrained small-scale modes are supplemented with random realizations in order to recover the full cosmological power spectrum.

The simulation was performed in a periodic box of
$500\,h^{-1}\mathrm{Mpc}$ on a side with $3072^3$ particles, initialized at $z=100$. The subsequent gravitational evolution from $z=100$ to $z=0$ was computed using a memory-optimized version of \texttt{GADGET-2} \citep{Springel2005}. Although the forward reconstruction stage relies on approximate dynamical models, the final cosmological evolution is fully non-linear and self-consistently generates dark matter haloes, filaments, and subhalo populations through the $N$-body gravitational solver.

The particle mass resolution is
$3.1 \times 10^8\,h^{-1}M_{\odot}$, sufficient to resolve the low-mass subhaloes used as tracers in our void analysis. Dark matter haloes and subhaloes are identified using a combination of the Friends-of-Friends (FOF) algorithm \citep{Davis1985} and the \texttt{SUBFIND} algorithm \citep{Springel2001}.

The reconstruction constraints are expected to be most reliable on large and intermediate scales, while residual uncertainties remain significant on highly non-linear small scales. 
Because the observational constraints are applied primarily at the massive-group scale with masses exceeding $10^{12}\,h^{-1}M_{\odot}$, the ELUCID simulation mainly constrains the large-scale structure of the cosmic web. Previous studies have shown that ELUCID reasonably reproduces prominent nearby structures such as the Coma cluster and the Sloan Great Wall \citep{WangHuiyuan2016, Chen2019, Luo2024}.

At smaller scales, however, ELUCID is not expected to provide a deterministic one-to-one positional reconstruction of individual galaxies or subhaloes. Instead, the low-mass subhalo population is generated statistically through non-linear gravitational evolution within the reconstructed large-scale environment. Consequently, the simulation should be interpreted as a statistically and environmentally consistent constrained realization rather than an exact reconstruction of the true local dark matter field.

In the following analysis, we utilize subhaloes extracted from both the reconstructed SDSS-like survey volume and the full periodic simulation box to investigate how tracer selection, survey geometry, and reconstruction procedures influence the statistical properties of cosmic voids.

\subsection{Algorithm for detecting voids}
\label{sec:algorithm}

To test the robustness of our void catalogues and to investigate how void properties depend on the chosen identification method, we use two different algorithms implemented in the \texttt{VAST} \citep{Douglass2023} package: \texttt{VoidFinder} and \texttt{REVOLVER}, the latter of which builds on the \texttt{ZOBOV} watershed technique. Although \texttt{REVOLVER} is specifically tailored for real-space coordinates and reconstructed density fields, it also supports multiple criteria for defining voids.

In our morphological analysis (Section~\ref{sec:void_shape}), we explore three distinct variants of the watershed framework, all implemented within the \texttt{VAST} software ecosystem. These include the \texttt{REVOLVER}-style single-zone selection \citep{Nadathur2019c}, \texttt{VIDE}-style zone linking \citep{Sutter2015}, and the standard \texttt{ZOBOV} statistical significance pruning \citep{Neyrinck2008}. By utilizing these unified implementations within \texttt{VAST}, rather than their respective original codebases, we ensure that the observed morphological differences arise solely from the underlying pruning logic rather than disparate numerical treatments.

To account for survey geometry and enable a consistent comparison between the simulation and observational catalogues, we apply the SDSS DR7 angular mask and radial selection function directly to the ELUCID subhalo distribution before performing void identification. This procedure constructs a survey-like mock catalogue with the same effective observational boundaries and selection effects as the SDSS sample, ensuring that the void-finding algorithms are applied under matched survey conditions.

In addition to the masked reconstruction volume, we also perform void identification on the full unmasked $(500~h^{-1}\mathrm{Mpc})^3$ periodic simulation box. This complementary analysis allows us to examine the influence of survey boundaries and finite-volume effects on the resulting void statistics.

The \texttt{VoidFinder} algorithm is a geometric approach that identifies voids by
maximizing inscribed spheres within underdense regions \citep{Hoyle2002}. It first populates these regions with spheres and selects non-overlapping maximal spheres, which are subsequently merged with smaller neighboring spheres that meet specific overlap criteria to define the final geometry. For \texttt{VoidFinder} voids, we restrict our analysis to ``interior'' voids that do not intersect the survey boundary by requiring the \textit{edge} flag provided in the \texttt{VAST} package to be zero. In contrast, watershed-based algorithms utilize a Voronoi tessellation technique to estimate the local density field and define voids as basins in this density landscape. For these voids, we categorize a candidate as an ``interior'' void only if its edge-to-total area ratio is less than 0.1, ensuring our analysis is not contaminated by boundary effects.

Although our methodology generally follows the criteria in \citet{Douglass2023}, there is a subtle but important difference in the treatment of the minimum void radius. During the identification process using the \texttt{VAST} package, we implement a lower initial search threshold of $r_{\mathrm{min}} = 5 \, h^{-1}\mathrm{Mpc}$, while only voids with an effective radius $r_{\mathrm{e}} \ge 10 \, h^{-1}\mathrm{Mpc}$ are retained for the final statistical analysis. This ``over-sampling'' approach ensures a higher degree of completeness near the $10 \, h^{-1}\mathrm{Mpc}$ boundary compared to a hard truncation at the initial search stage. Consequently, our void abundance at the small-radius end exhibits a slight enhancement relative to the results of \citet{Douglass2023}, providing a more continuous and robust distribution across the selection threshold.

Furthermore, to investigate the impact of mass-bias effects, we utilize the ELUCID simulation to construct four subhalo samples partitioned by their mass. It is important to emphasize that throughout this study, $M_{\mathrm{h}}$ specifically refers to the subhalo mass rather than the mass of the host halo. To ensure consistent statistical precision across different density regimes, we construct these four subsamples to contain an equal number of subhaloes, corresponding to the following mass ranges of $\log (M_{\mathrm{h}} / \text{M}_{\odot})$: $[10.9, 11.0)$, $[11.0, 11.2)$, $[11.2, 11.5)$, and $\ge 11.5$.

\section{Results}
\label{sec:results}

Using the void catalogues constructed from both the SDSS observations and the ELUCID simulations, we present a systematic analysis of the statistical properties of cosmic voids. To ensure robustness, we restrict our analysis to ``interior'' voids that are not truncated by survey or simulation boundaries, thereby minimizing potential boundary-induced biases.

We focus on three key aspects of void statistics: the size distribution, three-dimensional shape, and radial density profile. Throughout this section, we compare results across different data realizations (redshift space, real space, and reconstructed simulation), tracer selections, and void-finding algorithms,  with the aim of assessing the stability of void properties under these variations.

\subsection{Void size distribution}
\label{sec:size_distribution}

\begin{figure*}
\centering
\includegraphics[width=0.9\textwidth]{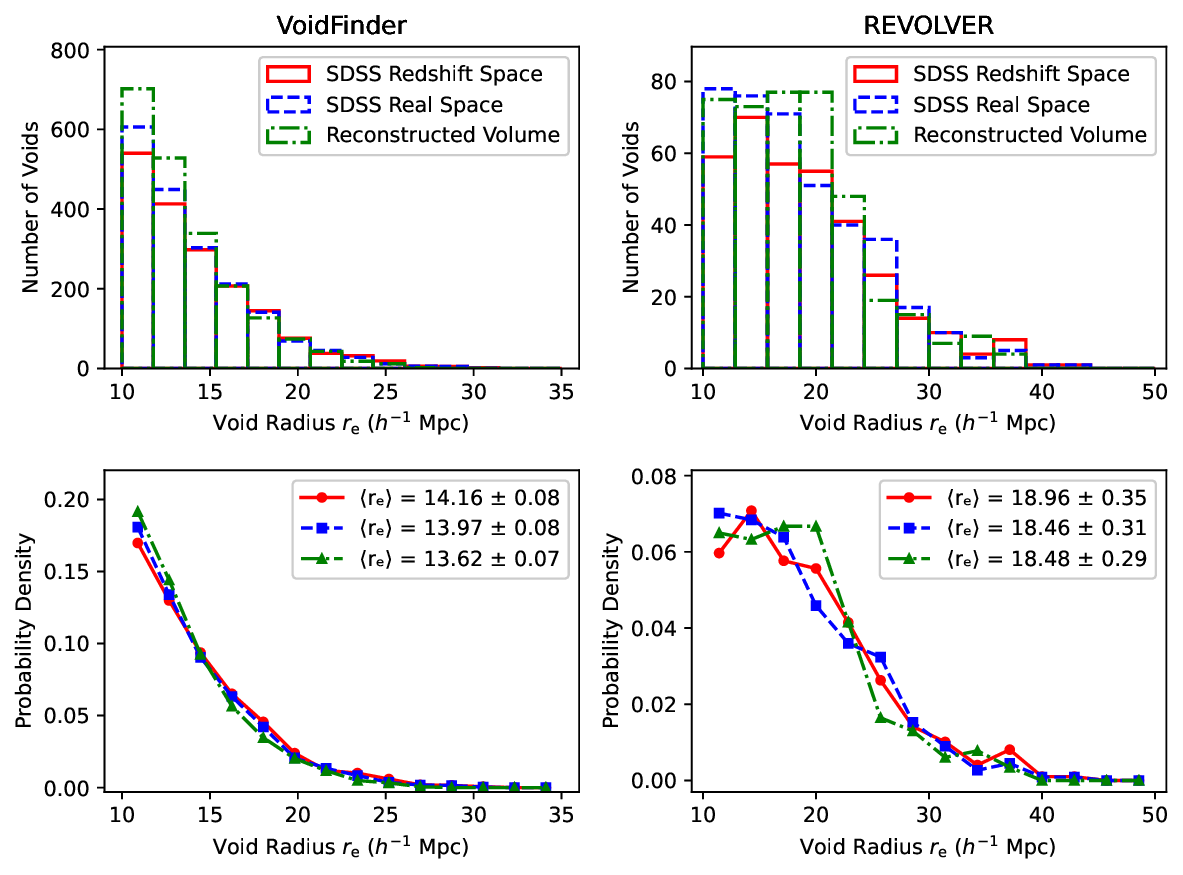}
\caption{Distribution of void effective radii $r_{\mathrm{e}}$ identified with \texttt{VoidFinder} (left) and \texttt{REVOLVER} (right) for SDSS DR7 galaxies (real and redshift space) and ELUCID simulation subhaloes in the reconstructed volume. The top panels show number counts, while the bottom panels show the corresponding probability density functions (PDFs). The mean radius $\langle r_{\rm e} \rangle$ for each sample is indicated in the bottom panels with its standard error of the mean.}
\label{fig:radius_obs}
\end{figure*}

\begin{figure*}
\centering
\includegraphics[width=0.9\textwidth]{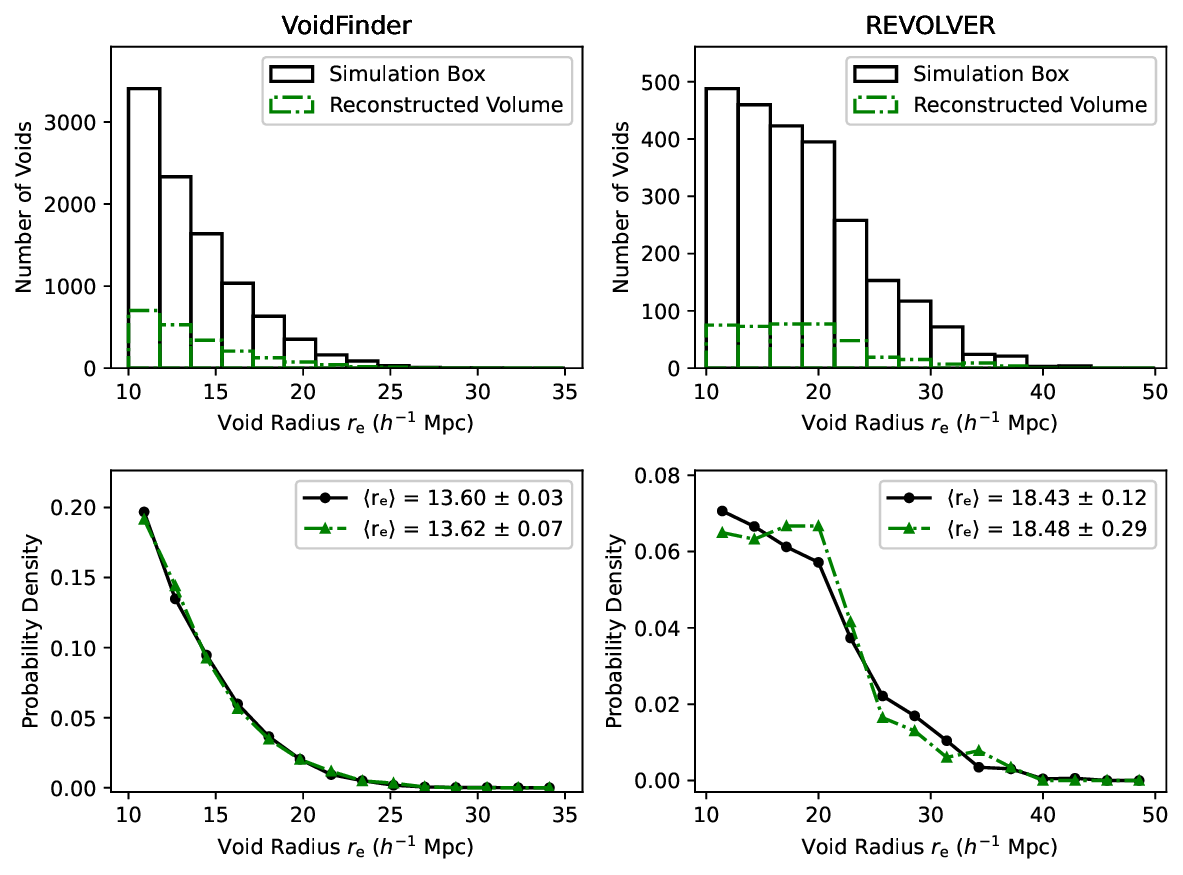}
\caption{Comparison of the PDFs of void effective radii measured in the full ELUCID simulation box and in the reconstructed volume. The reconstructed volume reproduces the geometry and volume of the observational sample, while the full simulation box represents the entire computational domain.}
\label{fig:box_rev}
\end{figure*}

\begin{figure*}
\centering
\includegraphics[width=0.9\textwidth]{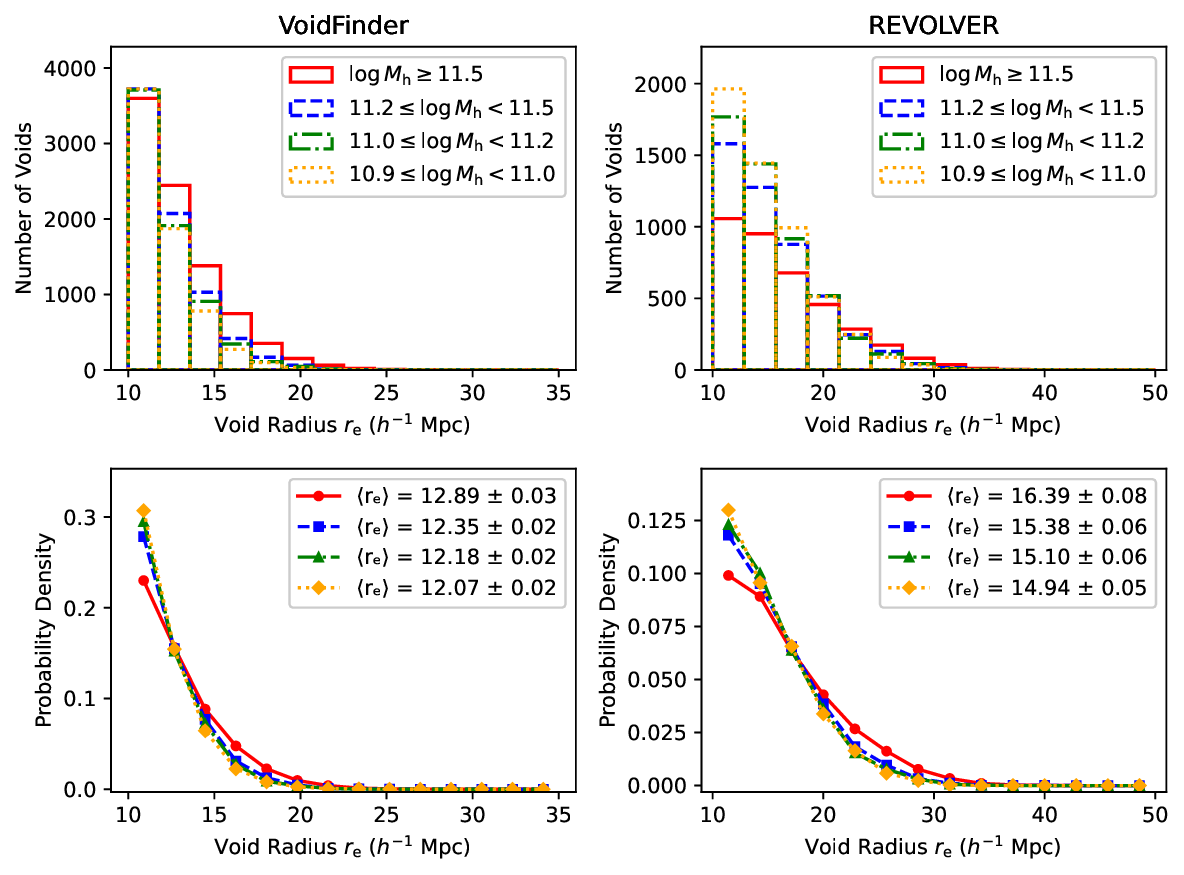}
\caption{Dependence of void effective radii on subhalo mass in the full ELUCID simulation box. The top panels show differential number counts, while the bottom panels show the PDFs of void effective radii $r_{\mathrm{e}}$. Left and right columns correspond to results from \texttt{VoidFinder} and \texttt{REVOLVER}, respectively. Four subhalo mass bins are shown (see legends), each containing an equal number of haloes.}
\label{fig:box_bins}
\end{figure*}

Void populations are commonly characterized using the void size function (VSF), defined as the number density of voids per unit radius interval (e.g., \citealt{Sheth2004, Jennings2013, Sutter2015, Contarini2019}). In this work, we instead present the distributions of void effective radii primarily in terms of differential number counts and normalized probability density functions (PDFs), which facilitate direct comparisons across different data realizations, tracer selections, and void-finding algorithms. We have verified that our main conclusions remain unchanged when using the conventional VSF.

\subsubsection{Comparison of observational and reconstructed voids}
\label{sec:void_comparison}

In this subsection, we compare the void populations identified in the SDSS DR7 galaxy sample with those obtained from the ELUCID simulation within the reconstructed survey volume. As discussed in Section~\ref{sec:elucid_sim}, the ELUCID simulation represents a dynamically reconstructed real-space realization constrained by the large-scale structure inferred from the SDSS galaxy-group catalogue after approximate RSD correction. Comparing the SDSS redshift-space sample, the reconstructed SDSS real-space sample, and the ELUCID realization therefore provides a useful framework for assessing how void statistics respond to observational RSD effects, reconstruction procedures, and void-finder methodologies.

Figure~\ref{fig:radius_obs} presents the distributions of void effective radii, $r_{\mathrm{e}}$, identified by the two algorithms across different data realizations. The top panels show the differential number counts, while the bottom panels display the corresponding PDFs. The SDSS redshift-space results are shown as red solid lines, the real-space results as blue dashed lines, and the ELUCID simulation reconstructed volume as green dash-dotted lines. The mean effective radius, $\langle r_{\mathrm{e}} \rangle$, and its associated statistical uncertainty are indicated in each PDF panel. Unless otherwise specified, all error bars presented in this study denote the standard error of the mean (SEM), defined as $\text{SEM} = \sigma / \sqrt{N_{\rm void}}$.

To ensure a fair comparison, all catalogues are analyzed within the same survey geometry defined by the SDSS DR7 mask. In addition, we match the tracer abundance by selecting subhaloes in the reconstructed volume with a mass threshold of $M_{\mathrm{h}} > 10^{11.8}~h^{-1}M_{\odot}$, such that the total number of tracers is consistent with that of the SDSS real-space sample. This procedure minimizes systematic differences between the catalogues.

We further restrict the analysis to ``interior'' voids with $r_{\mathrm{e}} > 10~h^{-1}\mathrm{Mpc}$ in order to suppress small-scale noise and boundary-related artifacts. A clear and systematic difference is observed between the two void-finding algorithms. \texttt{VoidFinder} identifies a large population of relatively small voids, with a distribution peaking near the imposed radius threshold and a mean radius of $\sim 14~h^{-1}\mathrm{Mpc}$. In contrast, \texttt{REVOLVER} yields fewer but significantly larger voids, with a broader distribution extending beyond $40~h^{-1}\mathrm{Mpc}$ and a mean radius of $\sim 18.5~h^{-1}\mathrm{Mpc}$. 

This discrepancy reflects the fundamental methodological differences between the two approaches. The sphere-based \texttt{VoidFinder} algorithm identifies maximal empty spheres and tends to partition extended underdense regions into multiple smaller voids. In contrast, the watershed-based \texttt{REVOLVER} algorithm merges adjacent low-density basins into larger coherent structures bounded by density ridges, resulting in fewer but larger voids.

The comparison between the SDSS and ELUCID catalogues further reveals that the level of agreement depends strongly on the void-finding methodology.
\begin{itemize}
\item 
For the watershed-based \texttt{REVOLVER} algorithm, the ELUCID realization and the reconstructed SDSS real-space sample exhibit broadly similar void size distributions and comparable characteristic scales. The corresponding PDFs show noticeable bin-to-bin fluctuations, which likely reflect the relatively small number of \texttt{REVOLVER} voids compared with the \texttt{VoidFinder} voids. Nevertheless, the overall distributions remain qualitatively consistent across the two datasets. The mean effective radius in ELUCID is
$\langle r_{\rm e} \rangle = 18.48 \pm 0.29~h^{-1}\mathrm{Mpc}$,
compared with
$\langle r_{\rm e} \rangle = 18.46 \pm 0.31~h^{-1}\mathrm{Mpc}$
for the SDSS reconstructed real-space sample. This close agreement in the characteristic radius suggests that the large-scale void population identified by \texttt{REVOLVER} is comparatively less sensitive to residual reconstruction uncertainties and is primarily governed by the large-scale geometry of the cosmic web. 

\item 
For the sphere-based \texttt{VoidFinder} algorithm, a somewhat larger offset is observed. The mean radius in ELUCID
($13.62 \pm 0.07~h^{-1}\mathrm{Mpc}$)
is slightly smaller than both the SDSS redshift-space
($14.16 \pm 0.08~h^{-1}\mathrm{Mpc}$)
and reconstructed real-space
($13.97 \pm 0.08~h^{-1}\mathrm{Mpc}$)
results. This behavior likely reflects the stronger sensitivity of maximal-empty-sphere methods to the precise spatial positions of individual tracers. Residual uncertainties associated with the reconstruction procedure, local tracer stochasticity, or small-scale positional differences between the simulation and observations can influence the sphere-expansion process and thereby modify the resulting void-radius distribution. 
\end{itemize}

Overall, the comparison indicates that the large-scale void population identified in the ELUCID constrained realization remains broadly consistent with the reconstructed SDSS real-space sample, particularly for the watershed-based \texttt{REVOLVER} catalogues. At the same time, the remaining differences between the datasets highlight that void statistics may also depend on reconstruction uncertainties, local tracer configurations, and algorithm-dependent sensitivities, especially for void definitions that rely strongly on small-scale geometric properties.

\subsubsection{Statistical consistency with the full simulation box}

To assess the statistical robustness of our results, we compare the void properties derived from the reconstructed volume with those obtained from the full $(500~h^{-1}\mathrm{Mpc})^3$ ELUCID simulation box. While the reconstructed volume is uniquely valuable due to its correspondence with the observed large-scale structure, it samples only a limited region of the full simulation and may therefore be affected by cosmic variance.

As shown in Figure~\ref{fig:box_rev}, the PDFs of void radii in the reconstructed volume closely match those from the full simulation box. For both void-finding algorithms, the mean effective radii are statistically consistent between the two samples. For example, \texttt{VoidFinder} yields $\langle r_{\mathrm{e}} \rangle = 13.60 \pm 0.03~h^{-1}\mathrm{Mpc}$ in the full box, consistent with $13.62 \pm 0.07~h^{-1}\mathrm{Mpc}$ in the reconstructed volume.

A more detailed comparison further supports this consistency. The median radius and interquartile range of the reconstructed sample agree closely with those of the full box. Moreover, the number of \texttt{VoidFinder} voids in the reconstructed volume ($2052$) corresponds to $\sim 21.2\%$ of the full-box total ($9692$), in good agreement with the geometric volume fraction of $\sim 21.77\%$. This near-linear scaling indicates that the reconstructed volume largely preserves the statistical homogeneity of the void population, with minor deviations attributable primarily to boundary exclusions.

For \texttt{REVOLVER}, the larger and more extended voids are more sensitive to boundary effects, leading to a somewhat lower count ratio ($\sim 16.7\%$). Nevertheless, the stability of the PDF shapes and the consistency of the mean radii ($\langle r_{\mathrm{e}} \rangle \approx 18.4~h^{-1}\mathrm{Mpc}$) confirm that the reconstructed volume provides a statistically representative sampling of the underlying density field. This validates the use of the full simulation box for subsequent analyses requiring higher statistical precision.

\subsubsection{Mass dependency of void radius distribution}
\label{sec:mass_abundance}

To investigate the impact of tracer bias, we divide the subhaloes in the full ELUCID simulation box into four mass bins, corresponding to the following mass ranges in $\log (M_{\mathrm{h}} / \msun)$: $[10.9, 11.0)$, $[11.0, 11.2)$, $[11.2, 11.5)$, and $\ge 11.5$. To isolate mass effects from number density variations, each bin contains an identical number of subhaloes ($N = 1{,}327{,}317$). Although this choice differs from the primary sample selection ($\log (M_{\mathrm{h}}/\msun) \ge 11.8$), it maximizes statistical power and enables a controlled exploration of mass-dependent trends.

Figure~\ref{fig:box_bins} shows the resulting void radius distributions. A clear monotonic dependence on subhalo mass is observed for both algorithms. For \texttt{VoidFinder}, the mean radius increases from $12.07 \pm 0.02~h^{-1}\mathrm{Mpc}$ to $12.89 \pm 0.03~h^{-1}\mathrm{Mpc}$ across the mass range. A similar but more pronounced trend is found for \texttt{REVOLVER}, with $\langle r_{\mathrm{e}} \rangle$ increasing from $14.94 \pm 0.05~h^{-1}\mathrm{Mpc}$ to $16.39 \pm 0.08~h^{-1}\mathrm{Mpc}$.

Interestingly, the two algorithms exhibit opposite trends in void abundance as a function of tracer mass. For \texttt{VoidFinder}, the total number of identified voids increases with increasing tracer mass, as more strongly biased tracers generate larger contiguous empty regions, allowing a greater number of maximal spheres to satisfy the void selection criteria. In contrast, \texttt{REVOLVER} shows a decreasing void count with increasing tracer mass, reflecting the merging of adjacent underdense regions as high-bias tracers sparsely sample the density ridges that would otherwise separate individual voids.

These results indicate that while higher-mass tracers delineate larger and more coherent underdense regions, the inferred void abundance depends sensitively on the void definition. This highlights the coupled role of tracer bias and algorithmic methodology in shaping void statistics, and underscores the importance of consistent definitions when comparing observations with simulations.

\subsection{Void sphericity and triaxiality}
\label{sec:void_shape}

\begin{table*}
\centering
\caption{Comparison of cosmic void shape parameters ($s$ and $T$) across different watershed-based algorithms and void catalogues. The sample includes voids with effective radii $r_{\rm e} > 10 \, h^{-1}$~Mpc and an edge-to-total area ratio below $0.1$. SEM denotes the standard error of the mean, calculated as $\sigma / \sqrt{N_{\rm void}}$.}
\label{tab:void_shapes}
\begin{tabular}{llccccccccc}
\toprule
Catalogue & Method & $N_{\rm void}$ & \multicolumn{2}{c}{Sphericity $s$ ($c/a$)} & \multicolumn{3}{c}{Percentiles ($s$)} & \multicolumn{2}{c}{Triaxiality $T$} & \multicolumn{1}{c}{Median $T$} \\
\cmidrule(lr){4-5} \cmidrule(lr){6-8} \cmidrule(lr){9-10}
& & & Mean $\pm$ SEM & Std & Q1 & Med & Q3 & Mean $\pm$ SEM & Std & Med \\
\midrule
\multirow{3}{*}{SDSS Redshift Space} 
& REVOLVER & 346 & $0.841 \pm 0.003$ & 0.056 & 0.807 & 0.849 & 0.880 & $0.309 \pm 0.011$ & 0.202 & 0.277 \\
& VIDE     & 340 & $0.838 \pm 0.003$ & 0.055 & 0.805 & 0.843 & 0.877 & $0.294 \pm 0.011$ & 0.199 & 0.269 \\
& ZOBOV    & 340 & $0.837 \pm 0.003$ & 0.055 & 0.804 & 0.843 & 0.877 & $0.294 \pm 0.011$ & 0.200 & 0.260 \\
\midrule
\multirow{3}{*}{SDSS Real Space} 
& REVOLVER & 389 & $0.844 \pm 0.003$ & 0.055 & 0.809 & 0.853 & 0.885 & $0.316 \pm 0.010$ & 0.207 & 0.261 \\
& VIDE     & 384 & $0.841 \pm 0.003$ & 0.054 & 0.803 & 0.849 & 0.881 & $0.309 \pm 0.011$ & 0.209 & 0.249 \\
& ZOBOV    & 383 & $0.841 \pm 0.003$ & 0.055 & 0.803 & 0.848 & 0.881 & $0.310 \pm 0.011$ & 0.211 & 0.244 \\
\midrule
\multirow{3}{*}{Reconstructed Volume} 
& REVOLVER & 404 & $0.850 \pm 0.002$ & 0.050 & 0.823 & 0.854 & 0.885 & $0.325 \pm 0.010$ & 0.199 & 0.297 \\
& VIDE     & 402 & $0.850 \pm 0.003$ & 0.051 & 0.822 & 0.855 & 0.885 & $0.321 \pm 0.010$ & 0.200 & 0.275 \\
& ZOBOV    & 398 & $0.850 \pm 0.003$ & 0.051 & 0.822 & 0.855 & 0.886 & $0.323 \pm 0.010$ & 0.202 & 0.276 \\
\bottomrule
\end{tabular}
\end{table*}

\begin{figure*}
\centering
\includegraphics[width=0.9\textwidth]{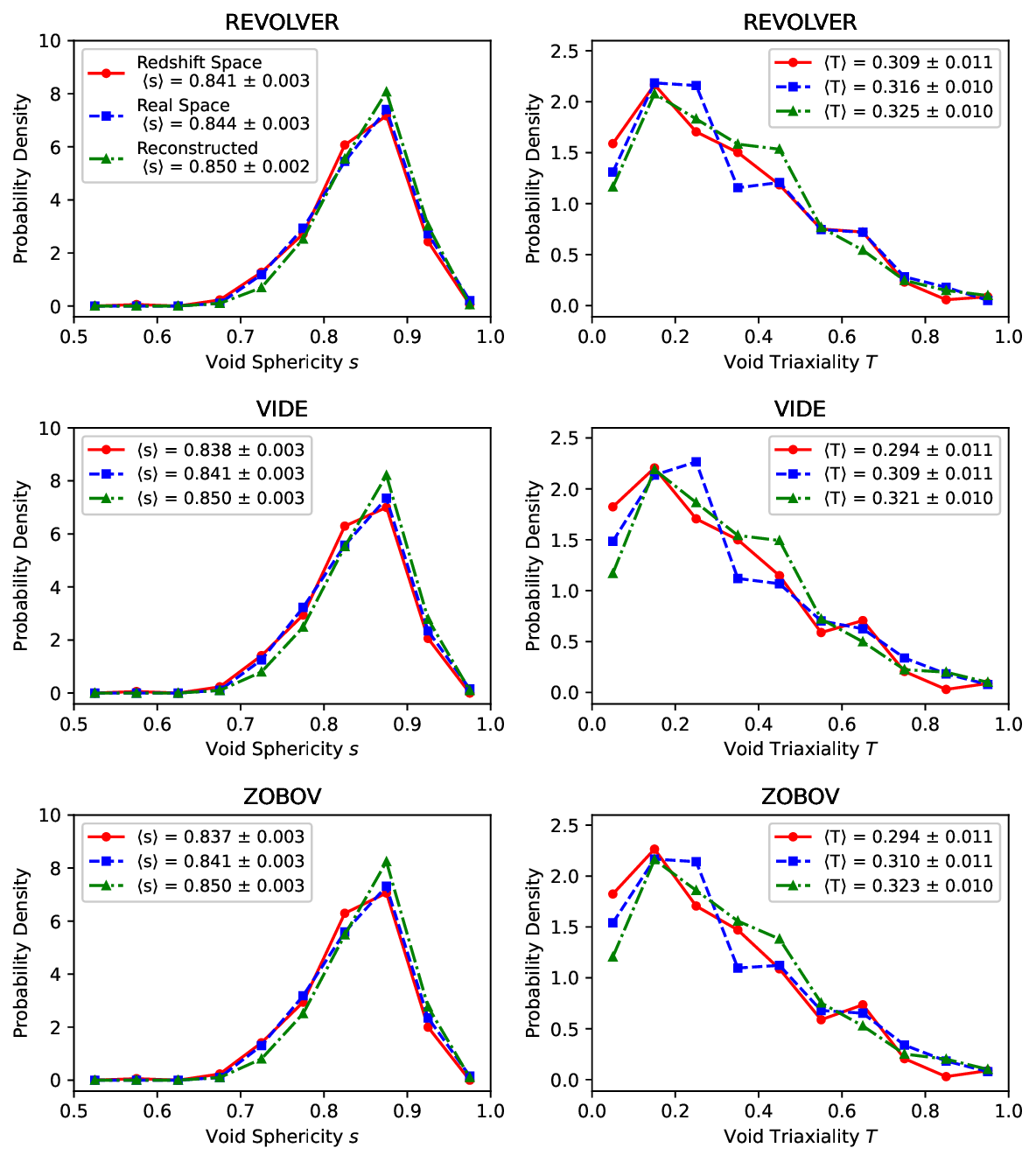}
\caption{Comparison of void sphericity ($s$, left panels) and triaxiality ($T$, right panels) measured with three different watershed-based algorithms. The rows from top to bottom correspond to the \texttt{REVOLVER}, \texttt{VIDE}, and \texttt{ZOBOV} methods, respectively. Within each panel, the distributions are compared across three samples: SDSS redshift space, SDSS real space, and the reconstructed volume, shown as solid, dashed, and dash-dotted lines. All voids satisfy a radius cut of $r_{\mathrm{e}} \geq 10\,h^{-1}\,\mathrm{Mpc}$.}
\label{fig:shape_obs}
\end{figure*}

\begin{figure*}
\centering
\includegraphics[width=0.9\textwidth]{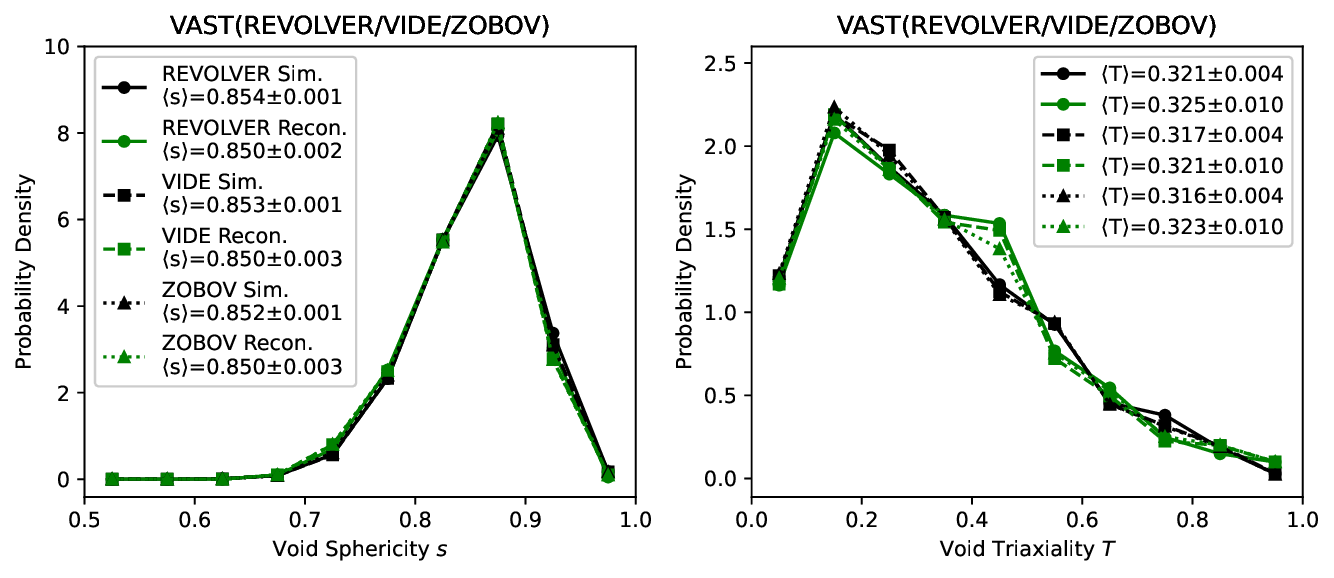}
\caption{Comparison of void sphericity ($s$, left panel) and triaxiality ($T$, right panel) between the full simulation box and the reconstructed volume in the ELUCID simulation. Results are calculated using \texttt{REVOLVER} (solid), \texttt{VIDE} (dashed), and \texttt{ZOBOV} (dotted) algorithms in the \texttt{VAST} package.
}
\label{fig:shape_box}
\end{figure*}

\begin{figure*}
\centering
\includegraphics[width=0.9\textwidth]{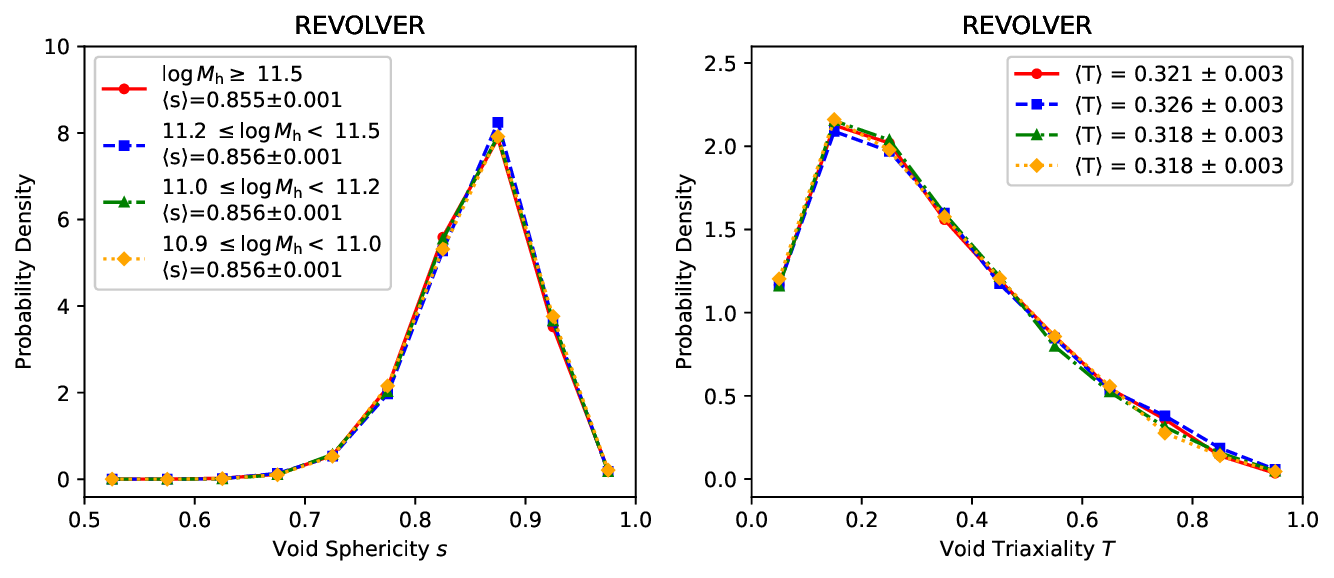}
\caption{PDFs of void sphericity ($s$, left panel) and triaxiality ($T$, right panel) across four distinct subhalo mass bins in the ELUCID simulation. All results are measured with the \texttt{REVOLVER} algorithm. The different subhalo populations are shown with distinct line styles and markers (see legends).}
\label{fig:shape_bins}
\end{figure*}

Beyond the size distribution, void morphology provides a more nuanced probe of the dynamical assembly of the cosmic web. By construction, \texttt{VoidFinder} identifies voids by maximizing inscribed spheres \citep{Hoyle2002}, a geometric prior that naturally favors spherical configurations. This design introduces a tendency toward spherical shapes, making it less suitable for accurately characterizing the intrinsic asphericity and triaxiality arising from the anisotropic evolution of large-scale structures. In contrast, watershed-based algorithms employ Voronoi-based density estimators that allow for flexible void boundaries, enabling them to better capture the complex deformations induced by external tidal fields. Therefore, in this subsection we focus exclusively on voids identified with the watershed-based algorithm. 

\subsubsection{Comparison of observational and reconstructed samples}

To ensure that our morphological conclusions are independent of specific void definitions, we compare results across the three pruning schemes previously described: \texttt{REVOLVER}-style, \texttt{VIDE}-style, and \texttt{ZOBOV} significance pruning. It is important to emphasize that all three variants are executed using the unified \texttt{VAST} implementation. This approach eliminates potential systematic biases arising from disparate numerical treatments in different software packages, ensuring that any observed differences in sphericity and triaxiality are purely consequences of the underlying watershed partitioning logic.

We quantify the three-dimensional morphology using the semi-axes $a, b,$ and $c$ ($a \ge b \ge c$) derived from the inertia tensor of the void's member tracers. The void shapes are then characterized by the sphericity $s$, defined as:
\begin{equation}
s = c/a,
\label{eq:sphericity}
\end{equation}
and the triaxiality parameter $T$:
\begin{equation}
T = \frac{a^2 - b^2}{a^2 - c^2},
\label{eq:triaxiality}
\end{equation}
where $s \in (0, 1]$, with $s=1$ representing a perfect sphere. The triaxiality parameter $T$ further distinguishes the specific nature of the asphericity: $T < 0.3$ represents oblate (pancake-like) shapes, while $T > 0.7$ indicates prolate (filament-like) geometries. Voids with $T \approx 0.5$ are considered maximally triaxial.

Figure~\ref{fig:shape_obs} displays the PDF distributions of these morphological parameters, with the corresponding quantitative statistics summarized in Table~\ref{tab:void_shapes}. The sphericity ($s$) distributions exhibit remarkable consistency across all three pruning schemes. Notably, in the reconstructed volume, all methods yield a nearly identical mean sphericity of $\langle s \rangle = 0.85$ with a narrow dispersion ($\sigma \approx 0.05$). This high degree of sphericity suggests that watershed-defined voids identified in the VAST framework are remarkably regular structures. The consistency across different pruning schemes further indicates that this result is robust against variations in the void definition.

Beyond simple sphericity, the distribution of the triaxiality parameter $T$ (Fig.~\ref{fig:shape_obs}, right panels) provides a more nuanced characterization of void geometry. The triaxiality distributions exhibit a prominent peak at $T < 0.3$, and as detailed in Table~\ref{tab:void_shapes}, the median triaxiality ($T_{\rm med} \approx 0.25$--$0.29$) remains consistently lower than the mean. This indicates a population strongly skewed toward oblate (pancake-like) geometries, consistent with the influence of the surrounding tidal field, which preferentially compresses void boundaries along directions perpendicular to nearby filaments and walls.

\subsubsection{Statistical consistency with the full simulation box}

To further validate the fidelity of our morphological recovery, we compare the shape distributions between the full simulation box (black curves) and the reconstructed volume (green curves). As shown in Figure~\ref{fig:shape_box}, the results exhibit close agreement across all three pruning schemes: \texttt{REVOLVER} (solid), \texttt{VIDE} (dashed), and \texttt{ZOBOV} (dotted).

For the sphericity $s$, both the full simulation box and reconstructed samples yield a mean value of $\langle s \rangle \approx 0.85$, with over $99.5\%$ of voids maintaining $s > 0.7$. The triaxiality parameter $T$ similarly shows consistent behavior, with a mean value of $\langle T \rangle \approx 0.32$. Our classification indicates that more than $50\%$ of the identified voids are oblate ($T < 0.3$), while only about $5\%$ exhibit prolate geometries ($T > 0.7$). The strong overlap between the black and green curves across all algorithms demonstrates that the reconstruction accurately preserves not only the spatial distribution of voids but also their intrinsic three-dimensional geometric properties.

\subsubsection{Mass dependency of void sphericity and triaxiality}

Building upon the four mass-selected subsamples defined in Section~\ref{sec:mass_abundance}, 
we now examine whether the tracer mass and its associated clustering bias imprint any systematic signatures on the void morphology. As previously noted, the identical number density across these bins allows us to isolate the physical impact of tracer mass from statistical artifacts related to shot noise.

For the sake of conciseness, we present the mass-dependence analysis using the \texttt{REVOLVER} scheme as a representative in Figure~\ref{fig:shape_bins}. We have explicitly verified that both the \texttt{VIDE} and \texttt{ZOBOV} pruning schemes yield nearly identical results, exhibiting the same remarkable stability across all four mass regimes. This cross-algorithm consistency reinforces the conclusion that the extracted void morphology is a robust physical property, independent of both the watershed implementation and the tracer mass.

Our analysis reveals a striking degree of mass-invariance in void geometry. As illustrated in Figure~\ref{fig:shape_bins}, the PDFs of sphericity $s$ are virtually indistinguishable across the four subsamples, with the mean values consistently centered around $\langle s \rangle \approx 0.86$. Quantitatively, over $99.3\%$ of voids across all mass regimes maintain a near-spherical geometry ($s > 0.7$), while highly elongated structures ($s \le 0.4$) are almost entirely absent. The triaxiality distributions are similarly robust, consistently peaking in the oblate regime ($T < 0.3$), which characterizes approximately $53\%$ of the total population. 

In summary, the void morphology shows a high stability across all tracer populations. The characteristic sphericity remains consistently at $\langle s \rangle \approx 0.86$, with the PDFs exhibiting nearly identical shapes across the four mass bins. This behavior indicates that the geometric properties of watershed-defined voids are largely insensitive to tracer mass, even for the most strongly biased subsamples. The robustness of this result supports the reliability of the extracted void catalogue for subsequent cosmological analyses.

\subsubsection{Methodological origin of the high measured sphericity}

The relatively high mean sphericity identified in our analysis ($\langle s \rangle \approx 0.85$--$0.86$) exceeds the values reported in many previous studies, which typically find $\langle s \rangle \approx 0.4$--$0.8$ depending on the tracer population and the void identification algorithm \citep[e.g.,][]{Bos2012, Tavasoli2013, Ghafour2025}. This difference is primarily methodological and can be traced to specific design choices in the \texttt{VAST/VIDE} framework. 

A secondary contribution arises from the use of tracer-based, rather than grid-based, estimators. Unlike voxelized approaches that uniformly sample the void volume, \texttt{VAST} constructs the inertia tensor from a discrete set of tracer particles within watershed-defined basins. In sparsely populated void interiors this leads to limited and nearly isotropic sampling of the density field, which can mildly bias the inferred shapes toward rounder configurations.

The dominant effect, however, originates from the mapping between the inertia tensor eigenvalues ($\lambda_i$) and the ellipsoidal semi-axes ($l_i$). Following the \texttt{VIDE} convention \citep{Sutter2015}, \texttt{VAST} adopts a fourth-root scaling, $l_i \propto \lambda_i^{1/4}$, rather than the quadratic relation $l_i \propto \lambda_i^{1/2}$ expected for homogeneous ellipsoids. This nonlinear transformation compresses the dynamic range of the eigenvalues and systematically reduces the contrast between the principal axes. Consequently, intrinsically elongated or flattened voids are mapped to more spherical configurations, leading to an upward shift in the measured axis ratio $s = c/a$.

To quantify the impact of this definition, we perform a controlled numerical experiment in which the fourth-root scaling is replaced by the standard quadratic relation, while keeping the void catalogue and all other analysis steps unchanged. Under this modification, the mean sphericity decreases from $\langle s \rangle \approx 0.85$ to $\langle s \rangle \approx 0.72$, bringing the values closer to those reported in previous studies \citep{Tavasoli2013}. The triaxiality parameter correspondingly shifts from $\langle T \rangle \approx 0.32$ to $\approx 0.35$, indicating a broader distribution of intrinsic shapes.

These results show that the elevated sphericity is largely driven by the adopted eigenvalue-to-axis mapping rather than by the underlying properties of the void population. Importantly, however, replacing the fourth-root scaling with the quadratic relation does not qualitatively alter the behaviour of the shape statistics across different samples. Although the absolute values of the shape parameters change, their relative variation between samples remains essentially unchanged, indicating that the inferred morphological trends are not sensitive to the specific eigenvalue-to-axis mapping adopted.

Overall, these algorithmic choices imply that the absolute sphericity values should be interpreted with caution, as they partly reflect properties of the shape estimator itself. Nevertheless, because these procedures are applied consistently across all samples, the relative trends in void morphology—particularly their weak dependence on redshift-space distortions and tracer bias—remain robust.

\subsection{Void density profile}

\begin{table*}
\centering
\caption{Best-fit parameters of the re-parameterized Hamaus density profile for cosmic voids identified by \texttt{REVOLVER} and \texttt{VoidFinder}. The profiles are fitted using the functional form $\rho(r)/\rho_{\rm b} = 1 + \delta_c [1 - (x/f_s)^\alpha] / (1 + x^\beta)$, where $x = r/r_{\rm e}$. $N_{\rm void}$ denotes the number of interior voids used in the stacking, and $\langle r_{\rm e} \rangle$ is the mean effective radius in units of $h^{-1}\,\mathrm{Mpc}$. 
MAE represents the mean absolute residual measured in the transition region ($0.8 < r/r_{\rm e} < 1.2$).}
\label{tab:fit_params}
\begin{tabular}{llccccccc}
\toprule
catalogue & Method & $N_{\rm void}$ & $\langle r_{\rm e} \rangle$ & $\delta_c$ & $f_s$ & $\alpha$ & $\beta$ & MAE \\
\midrule
\multirow{2}{*}{SDSS Redshift Space} 
& \texttt{REVOLVER}   & 346  & 18.96 & $-0.815 \pm 0.022$ & $0.775 \pm 0.007$ & $2.986 \pm 0.161$ & $7.511 \pm 0.280$ & 0.039 \\
& \texttt{VoidFinder} & 1781 & 14.16 & $-0.933 \pm 0.004$ & $0.876 \pm 0.002$ & $5.056 \pm 0.085$ & $11.548 \pm 0.221$ & 0.197 \\
\midrule
\multirow{2}{*}{SDSS Real Space} 
& \texttt{REVOLVER}   & 389  & 18.46 & $-0.835 \pm 0.021$ & $0.774 \pm 0.007$ & $2.896 \pm 0.157$ & $7.485 \pm 0.275$ & 0.037 \\
& \texttt{VoidFinder} & 1877 & 13.97 & $-0.938 \pm 0.004$ & $0.871 \pm 0.002$ & $5.231 \pm 0.085$ & $12.311 \pm 0.227$ & 0.216 \\
\midrule
\multirow{2}{*}{Reconstructed Volume} 
& \texttt{REVOLVER}   & 404  & 18.48 & $-0.828 \pm 0.018$ & $0.759 \pm 0.007$ & $2.535 \pm 0.121$ & $6.956 \pm 0.234$ & 0.022 \\
& \texttt{VoidFinder} & 2052 & 13.62 & $-0.932 \pm 0.004$ & $0.893 \pm 0.002$ & $4.701 \pm 0.082$ & $12.012 \pm 0.257$ & 0.195 \\
\bottomrule
\end{tabular}
\end{table*}

\begin{figure}
\centering
\includegraphics[width=0.45\textwidth]{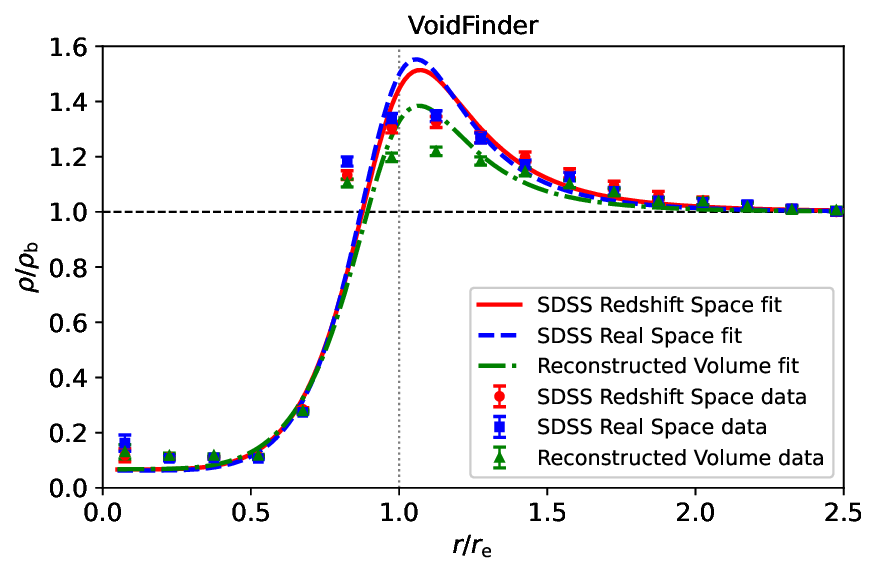}
\includegraphics[width=0.45\textwidth]{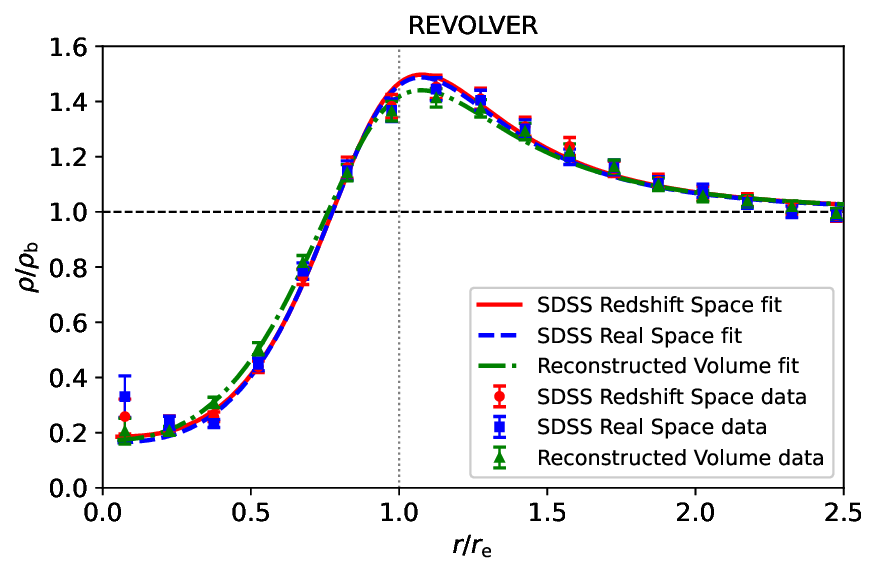}
\caption{Stacked radial number density profiles of cosmic voids identified by the \texttt{VoidFinder} (upper panel) and \texttt{REVOLVER} (bottom panel) algorithms. 
Markers represent the measured density ratios from the SDSS redshift-space (circles), SDSS real-space (squares), and the ELUCID reconstructed volume (triangles). 
The corresponding lines (solid, dashed, and dash-dotted) denote the best-fitting profiles obtained using the re-parameterized functional form (Eq.~\ref{eqn:vdf}) based on \citet{Hamaus2014}. 
The horizontal dashed and vertical dotted lines represent the background density $\rho/\rho_{\mathrm{b}} = 1$ and the effective void radius $r/r_{\mathrm{e}} = 1$, respectively.}
\label{fig:profile_obs}
\end{figure}

\begin{figure}
\centering
\includegraphics[width=0.45\textwidth]{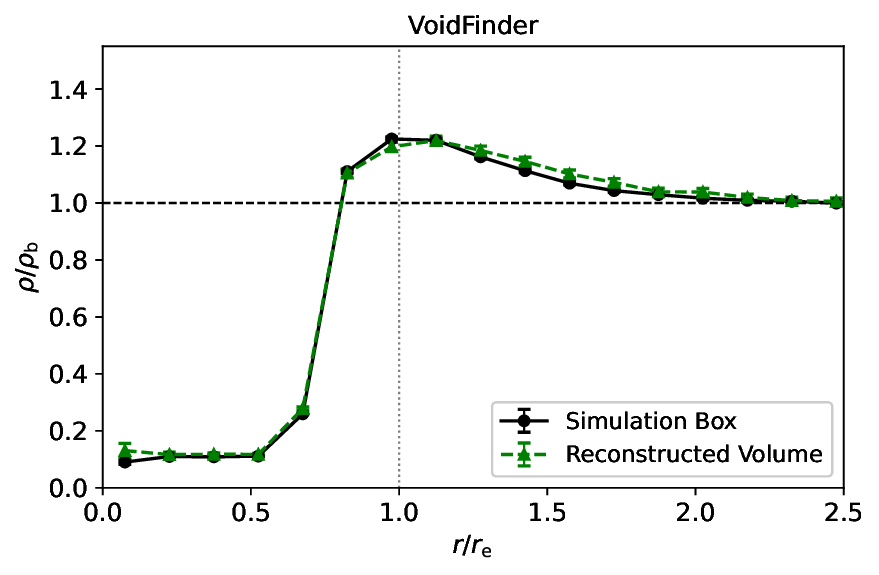}
\includegraphics[width=0.45\textwidth]{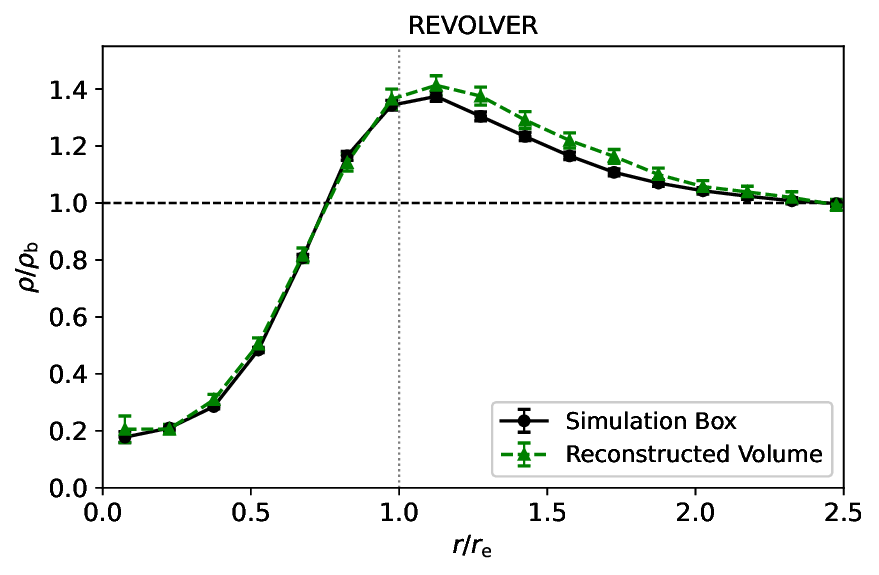}
\caption{Comparison of the stacked radial number density profiles of cosmic voids between the full simulation box and the reconstructed volume in the 
ELUCID simulation. The upper panel shows the results for \texttt{VoidFinder}, and the bottom panel for \texttt{REVOLVER}.}
\label{fig:profile_box}
\end{figure}

\begin{figure}
\centering
\includegraphics[width=0.45\textwidth]{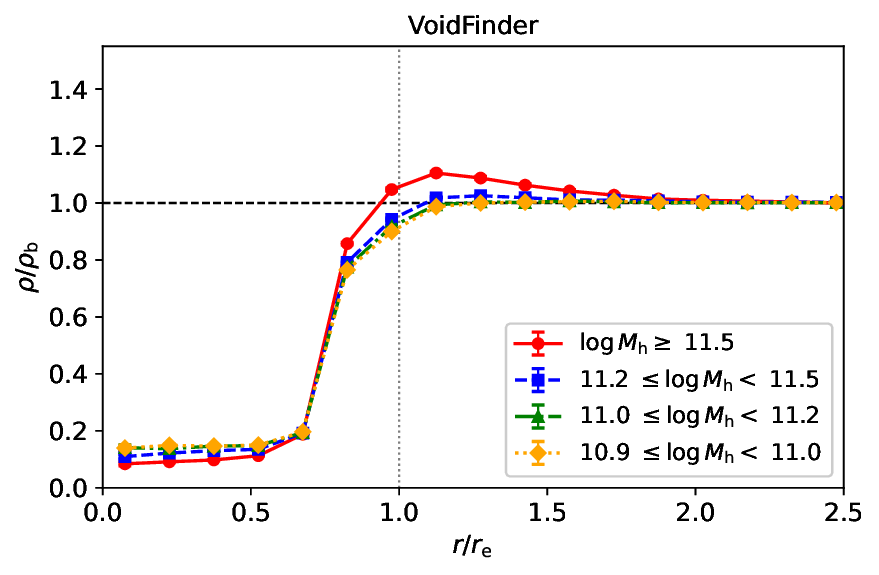}
\includegraphics[width=0.45\textwidth]{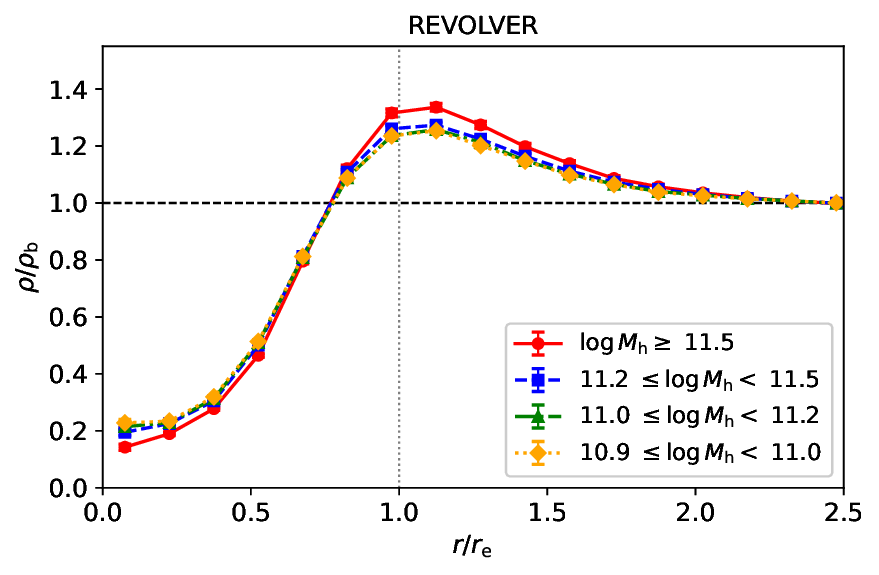}
\caption{Radial density profiles of voids identified from four different subhalo mass bins in the ELUCID simulation. The upper panel shows the results for \texttt{VoidFinder}, and the bottom panel for \texttt{REVOLVER}. Different colors represent different subhalo mass ranges as indicated in the legend. The horizontal dashed and vertical dotted lines represent $\rho/\rho_{\mathrm{b}} = 1$ and $r/r_{\mathrm{e}} = 1$, respectively.}
\label{fig:profile_bins}
\end{figure}

In this section, we investigate the stacked radial number density profiles $\rho(r)$ of the identified voids. To enable a direct comparison across different galaxy catalogues and void-finding algorithms, the local density is normalized by a sample-specific background density $\rho_{\mathrm{b}}$. We define $\rho_{\mathrm{b}}$ as the mean number density measured in the outer shells of the voids ($2.0 < r/r_{\mathrm{e}} < 3.0$). This normalization accounts for small variations in the global tracer density across samples—for example, the $\sim7\%$ lower density in the reconstructed volume relative to the full simulation box—likely reflecting large-scale density variations within the constrained region. With this definition, the resulting profiles emphasize differences in the internal structure of voids rather than offsets in the overall tracer density.

\subsubsection{Comparison of observational and reconstructed profiles}

To quantitatively characterize the internal structure of cosmic voids, we fit the stacked radial density profiles using a re-parameterized functional form based on \citet{Hamaus2014}:
\begin{equation}
\label{eqn:vdf}
\frac{\rho(r)}{\rho_{\mathrm{b}}} = 1 + \delta_c \frac{1 - (x/f_s)^\alpha}{1 + x^\beta},
\end{equation}
where $x = r/r_{\mathrm{e}}$ is the normalized radial distance from the void center. In this formulation, $\delta_c$ is a fitting parameter that characterizes the depth of the central underdensity in the stacked profile, while $\alpha$ and $\beta$ control the slopes of the inner and outer regions, respectively.

A key advantage of this re-parameterization is the explicit introduction of a physically interpretable transition scale $f_s$. By construction, the model satisfies $\rho/\rho_{\mathrm{b}} = 1$ at $x = f_s$. This differs from the original formulation of \citet{Hamaus2014}, where the transition radius is implicitly determined. The use of the normalized coordinate $x$ further enables a consistent comparison across void samples with different characteristic sizes.

Figure~\ref{fig:profile_obs} shows the stacked density profiles for voids identified by \texttt{VoidFinder} (upper panel) and \texttt{REVOLVER} (lower panel). All profiles broadly exhibit the canonical void density structure, characterized by a strongly underdense core followed by a compensation wall peaking at $r/r_{\mathrm{e}} \approx 1.1$. The best-fit parameters are summarized in Table~\ref{tab:fit_params}. A clear contrast emerges between the two void-finding algorithms: \texttt{REVOLVER} selects a sample of larger, dynamically more coherent voids for which the model provides an excellent description, achieving mean absolute errors ($\mathrm{MAE}$) as low as $\sim 0.02$. In contrast, \texttt{VoidFinder} identifies a larger population of smaller voids, but the resulting profiles show significant deviations from the fitted model, with $\mathrm{MAE} \sim 0.2$.

For the \texttt{REVOLVER} profiles, the fitted transition scale $f_s$ agrees closely with the transition radius independently estimated via direct interpolation, both yielding $r/r_{\mathrm{e}} \approx 0.76$--$0.77$. This agreement indicates that the functional form in Eq.~(\ref{eqn:vdf}) correctly captures the physically meaningful crossing point for watershed-based voids. Furthermore, the best-fit parameters obtained from the ELUCID reconstructed volume closely match those derived from the SDSS observational catalogues, indicating that the reconstructed ELUCID realization reproduces the overall structural characteristics of the observed void population at a statistically consistent level.

However, for \texttt{VoidFinder} profiles, systematic residuals are observed in the transition region ($0.8 < r/r_{\rm e} < 1.2$). The transition radius estimated via direct interpolation is $r/r_{\mathrm{e}} \approx 0.80$, whereas the fitted $f_s$ in Eq.~(\ref{eqn:vdf}) is systematically larger ($0.87$--$0.89$). To investigate this discrepancy, we tested a modified version of Eq.~(\ref{eqn:vdf}) by replacing the $x^\beta$ term in the denominator with $(x/f_s)^\beta$, effectively rescaling the entire profile relative to $f_s$. For \texttt{VoidFinder}, this adjustment yields a more consistent $f_s \approx 0.83$ and reduces the residuals to $\mathrm{MAE} \lesssim 0.08$. 

Nevertheless, for the \texttt{REVOLVER} profiles, this same modification significantly degrades the fitting performance, with the MAE increasing from $\sim 0.03$ to $\sim 0.10$ and introducing strong parameter degeneracies. In contrast, the fitting formulation of Eq.~(\ref{eqn:vdf}) provides an excellent description of the \texttt{REVOLVER} sample. This high level of agreement is expected, as both the \texttt{REVOLVER} algorithm and the original \citet{Hamaus2014} profile were developed within the framework of watershed-based void identification. Notably, our adopted form in Eq.~(\ref{eqn:vdf}) maintains strict mathematical consistency with the Hamaus universal profile in terms of its radial scaling: the denominator term $1+x^\beta$ preserves the standard normalization relative to the effective radius $r_{\rm e}$, while the numerator term $(x/f_s)^\alpha$ is physically equivalent to $(r/r_s)^\alpha$, where $r_s = f_s \cdot r_{\rm e}$ is the fitting transition radius. Therefore, adhering to the original scaling in Eq.~(\ref{eqn:vdf})—rather than shifting the denominator normalization to $f_s$—provides a more physically motivated and internally consistent benchmark for characterizing the structural diversity across different void populations.

This comparison demonstrates that the widely used Hamaus universal profile provides a physically consistent description primarily for watershed-defined voids, while its applicability to geometrically defined voids such as VoidFinder is more limited.

Beyond the fitting performance, we also examine the impact of ELUCID reconstruction on the detailed shape of the density profiles. In the reconstructed volume, the amplitude of the compensation wall exhibits a noticeable suppression relative to the observational samples. This effect is likely driven by the small-scale smoothing inherent in the ELUCID reconstruction process, which tends to reduce the density contrast of the filamentary and wall-like structures that define void boundaries. Notably, this suppression has a stronger impact on \texttt{VoidFinder} profiles, which are sensitive to small-scale variations in the discrete galaxy distribution. In contrast, the \texttt{REVOLVER} profiles remain remarkably stable across the different data sets. As a variant of the watershed-based \texttt{ZOBOV} method, \texttt{REVOLVER} identifies voids as density basins within the Voronoi tessellation; it depends primarily on the overall topology of the density field, which proves to be more robust against the moderate smoothing effects of the reconstruction.

Overall, the broad agreement between the observational and reconstructed profiles suggests that the ELUCID constrained realization captures the main structural properties of the observed void population, while residual differences likely reflect reconstruction uncertainties, tracer discreteness, and algorithm-dependent sensitivities.

\subsubsection{Statistical robustness and verification}

We further assess the robustness of the reconstruction by comparing the ELUCID volume with the full simulation box, as shown in Figure~\ref{fig:profile_box}. For both void-finding algorithms, the density profiles exhibit consistent structural features, including the underdense cores and well-defined transition radii.

In the upper panel, the \texttt{VoidFinder} profiles show a high degree of agreement across the entire radial range, indicating that the internal density structure of VoidFinder voids is well preserved in the reconstruction. In the lower panel, the \texttt{REVOLVER} profiles are also broadly consistent with those from the simulation box. The core underdensity and transition radius are accurately recovered, although the reconstructed volume exhibits a slightly elevated density ratio at larger radii ($r/r_{\mathrm{e}} \approx 1.5$).

This small offset likely reflects residual differences in large-scale clustering between the ELUCID volume and the full simulation box, rather than a systematic failure of the reconstruction. Overall, the close agreement between the two suggests that the ELUCID reconstruction preserves the main statistical and structural properties of cosmic voids at a broadly consistent level.

\subsubsection{Mass dependency of void density profiles}
\label{sec:mass_profile}

We investigate the dependence of void radial density profiles on the mass of the subhaloes used as tracers. Figure~\ref{fig:profile_bins} shows the stacked profiles for the four subhalo mass bins defined in Section~\ref{sec:mass_abundance}.

Both \texttt{REVOLVER} and \texttt{VoidFinder} reveal a consistent trend with tracer mass. For the three lower-mass bins ($10.9 \le \log M_{\mathrm{h}} < 11.5$), the profiles are nearly invariant and show substantial overlap. In contrast, the highest-mass bin ($\log M_{\mathrm{h}} \ge 11.5$) exhibits a clear enhancement of the density contrast, characterized by a deeper central depletion ($r/r_{\mathrm{e}} < 0.2$) and a higher compensation wall. The consistency of this behavior across both algorithms indicates that tracer bias plays a dominant role: more massive subhaloes are more efficiently evacuated from void interiors and more strongly clustered at the boundaries.

The primary difference between the two algorithms lies in the stability of the characteristic scales. For \texttt{REVOLVER}, the transition radius estimated via direct interpolation
remains nearly constant at $r/r_{\mathrm{e}} \approx 0.77$--$0.78$ across all mass bins. In contrast, \texttt{VoidFinder} exhibits a stronger mass dependence, with the transition radius shifting systematically from $r/r_{\mathrm{e}} \approx 0.89$ to $r/r_{\mathrm{e}} \approx 1.27$ toward lower tracer masses. 

Despite this variation in scale, both methods consistently capture the same mass-dependent enhancement of the density contrast in the highest-mass bin. This agreement indicates that the impact of tracer bias on void density profiles is robustly reflected by both algorithms. However, the transition radius derived from the REVOLVER catalogues remains nearly invariant across all mass bins, suggesting that the dynamical and topological definition of watershed voids leads to a more stable geometric scaling of the stacked profiles.

\section{Summary}
\label{sec_summary}

In this work, we have carried out a systematic investigation of the statistical properties of cosmic voids using the SDSS DR7 galaxy sample and the ELUCID constrained simulation. By combining geometry-based and watershed-based void-finding algorithms within the \texttt{VAST} framework, and by comparing void catalogues across redshift space, reconstructed real space, the ELUCID reconstructed volume, and the full simulation box, we assess the robustness of void statistics against key systematic effects, including reconstruction uncertainties, tracer bias, and algorithmic definitions.

Our main findings can be summarized as follows:

\begin{enumerate}

\item \textbf{Void shapes are relatively stable.}
The three-dimensional morphology of voids, characterized by sphericity $s$ and triaxiality $T$, remains stable across all data realizations and tracer selections. The mean sphericity consistently stays at $\langle s \rangle \approx 0.85\text{--}0.86$, with a dominant population of oblate voids ($T < 0.3$). This robustness suggests that void shapes are primarily linked to the large-scale gravitational environment and are comparatively insensitive to observational systematics and tracer-dependent effects.

\item \textbf{Void size statistics depend strongly on the identification algorithm.}
While qualitative trends are consistent, we identify systematic differences between void-finding methods. Watershed-based algorithms systematically produce larger voids with more pronounced compensation walls compared to geometry-based \texttt{VoidFinder}. These discrepancies arise from the fundamentally different definitions of void boundaries (basin merging versus maximal sphere packing), highlighting that void size-related statistics are intrinsically method-dependent.

\item \textbf{Tracer bias affects density profiles but not geometry.}
By exploiting the full ELUCID simulation box, we show that void properties exhibit a conditional dependence on tracer mass. Noticeable variations in the density profiles—specifically deeper central underdensities and stronger compensation walls—emerge only for the most massive subhaloes ($>10^{11.5}~h^{-1}M_{\odot}$). In contrast, void shapes remain nearly invariant across all mass bins, reinforcing their status as robust tracers of the underlying large-scale matter distribution.

\item \textbf{Constrained simulations provide a useful bridge between observations and theory.}
The broad agreement in void statistics between SDSS observations, the ELUCID reconstructed volume, and the full simulation box suggests that the ELUCID density field broadly reproduces the large-scale topology of the observed cosmic web. This consistency supports the use of constrained simulations as a valuable framework for studying how reconstruction uncertainties, tracer-dependent effects, and void-finding methodologies influence void statistics in the low-density Universe.

\end{enumerate}

Overall, our results demonstrate that not all void statistics are equally robust: geometric properties, particularly void shapes, are largely insensitive to observational and methodological uncertainties, whereas size-related quantities and density profiles exhibit a stronger dependence on void definitions and tracer properties. This distinction has important implications for the use of cosmic voids as cosmological probes.

Building on this work, the methodology and validation framework developed here provide a foundation for next-generation analyses based on the \texttt{ELUCID-DESI} project. With the increased depth and volume of DESI-BGS data, future studies will enable more precise investigations of void dynamics and the growth of large-scale structure, further advancing the role of cosmic voids in precision cosmology.

\section*{Acknowledgements}

This research is funded by various grants, including the National Natural Science Foundation of China (Nos. 12273088, 12595313, 12103037), the National SKA Program of China (grant No. 2025SKA0150100), and the National Key R\&D Programme of China (2023YFA1607800, 2023YFA1607804). Additional support comes from the CSST project (Nos. CMS-CSST-2021-A02, CMS-CSST-2025-A04), the CAS Project for Young Scientists in Basic Research (No. YSBR-092), Fundamental Research Funds for Central Universities, the 111 project (No. B20019) and the Shanghai Natural Science Foundation (grant No.19ZR1466800, 23JC1410200, ZJ20223-ZD-003). PW acknowledge financial support by the NSFC (No. 12473009), and also sponsored by Shanghai Rising-Star Program (No.24QA2711100). FS acknowledges the support from the State Key Laboratory of Dark Matter Physics and the Young Data Scientist Program of the China National Astronomical Data Center (No.NADC2025YDS-01).

This work is also supported by the High-Performance Computing Resource in the Core Facility for Advanced Research Computing at Shanghai Astronomical Observatory.

This work used {\tt Astropy:} a community-developed core Python package
and an ecosystem of tools and resources for astronomy \citep{Astropy2013, 
Astropy2018, Astropy2022}. 

Funding for the Sloan Digital Sky Survey IV has been provided by the
Alfred P. Sloan Foundation, the U.S. Department of Energy Office of
Science, and the Participating Institutions. SDSS acknowledges support
and resources from the Centre for High-Performance Computing at the
University of Utah. The SDSS website is www.sdss.org.

SDSS is managed by the Astrophysical Research Consortium for the
Participating Institutions of the SDSS Collaboration including the
Brazilian Participation Group, the Carnegie Institution for Science,
Carnegie Mellon University, the Chilean Participation Group, the
French Participation Group, Harvard-Smithsonian Center for
Astrophysics, Instituto de Astrof{\'i}sica de Canarias, The Johns
Hopkins University, Kavli Institute for the Physics and Mathematics of
the Universe (IPMU)/University of Tokyo, Lawrence Berkeley National
Laboratory, Leibniz Institut f{\"u}r Astrophysik Potsdam (AIP),
Max-Planck-Institut f{\"u}r Astronomie (MPIA Heidelberg),
Max-Planck-Institut f{\"u}r Astrophysik (MPA Garching),
Max-Planck-Institut f{\"u}r Extraterrestrische Physik (MPE), National
Astronomical Observatories of China, New Mexico State University, New
York University, University of Notre Dame, Observat{\'o}rio Nacional/
MCTI, The Ohio State University, Pennsylvania State University,
Shanghai Astronomical Observatory, United Kingdom Participation Group,
Universidad Nacional Aut{\'o}noma de M{\'e}xico, University of
Arizona, University of Colorado Boulder, University of Oxford,
University of Portsmouth, University of Utah, University of Virginia,
University of Washington, University of Wisconsin, Vanderbilt
University, and Yale University.

\section*{Data availability}
The data underlying this article will be shared on reasonable request with
the corresponding author.

\bibliographystyle{mnras}
\bibliography{bibliography}

\bsp    
\label{lastpage}
\end{document}